\documentclass[aps,pre,twocolumn,groupedaddress,showpacs,superscriptaddress]{revtex4}
\usepackage{graphicx}
\begin{document}

\title{Nonequilibrium Green's function method for thermal transport in junctions}

\author{Jian-Sheng Wang} 
\homepage[]{http://staff.science.nus.edu.sg/~phywjs/}
\affiliation{Center for Computational Science and Engineering, and
Department of Physics, National University of Singapore, Singapore
117542} 
\affiliation{Institute of High Performance Computing, 1 Science Park
Road, Singapore 117528}
\affiliation{Singapore-MIT Alliance, 4 Engineering Drive 3, Singapore
117576}
\author{Nan Zeng} 
\author{Jian Wang}
\affiliation{Center for Computational Science and Engineering, and
Department of Physics, National University of Singapore, Singapore
117542}
\author{Chee Kwan Gan}
\affiliation{Institute of High Performance Computing, 1 Science Park
Road, Singapore 117528}

\date{9 January 2007}

\begin{abstract}
We present a detailed treatment of the nonequilibrium Green's function
method for thermal transport due to atomic vibrations in
nanostructures.  Some of the key equations, such as self-energy and
conductance with nonlinear effect, are derived.  A self-consistent
mean-field theory is proposed.  Computational procedures are
discussed.  The method is applied to a number of systems including
one-dimensional chains, a benzene ring junction, and carbon nanotubes.
Mean-field calculations of the Fermi-Pasta-Ulam model are compared
with classical molecular dynamics simulations.  We find that
nonlinearity suppresses thermal transport even at moderately high
temperatures.
\end{abstract}


\maketitle


\section{Introduction}
Fourier's law describes the transport of heat in a macroscopic object.
The calculation of the thermal conductivity that appears in the
Fourier's law is fundamental and important in understanding the
properties of materials.  Such a calculation must be grounded upon
quantum mechanics with phonons as basic quasi-particles.  This is a
non-trivial task and was successfully carried out many years ago by
Peierls \cite{peierls,heat-transport-textbooks} using the Boltzmann
equation for phonons.  Molecular dynamic (MD) simulation is another
approach.  However, MD results are not correct at low temperatures as
it is purely a classical approach.

The early works are mostly concerned with bulk materials with many
degrees of freedom and periodic lattices \cite{rmp-review}.  In recent
years, more attentions are paid to heat transport in small or
low-dimensional systems such as carbon nanotubes and molecular
junctions \cite{cahill,galperin}.  In nano- to meso-scales, new
features come in.  In most of these conditions, the discreteness of
the atom is important.  The concept of phonons developed for the
periodic lattices is somewhat difficult to apply if a system does not
possess translational invariance, such as a nano-junction.  In such
situation, the Boltzmann-Peierls' approach is not applicable.

There are several alternative methods that have overcome the above
problems.  First, the Landauer formula \cite{landauer-thermal} is a
simple and clear description of the heat transport in purely ballistic
regime at low temperatures.  There are also a few other approaches
\cite{michel,haanggi} with different approximations such as the
description based on density matrices.  One of the features of the
existing theories is that they work in either the ballistic regime or
diffusive regime, but not both.  It is rather difficult to have a
complete theory that can encompass both regimes, apart from
phenomenological treatments \cite{chenG,canala}.

The nonequilibrium Green's function formalism described in this
article is a serious attempt to be such a complete theory.  This
theory is certainly ``first principles'' given the atomic potentials.
Several similar formulations have already been done at the elastic
level without nonlinear interactions
\cite{ciraci,mingo,yamamoto,dhar}.  In Ref.\cite{PRB-green}, we
presented a new formulation with the nonlinear interaction treated
systematically.  Mingo also gave a similar result
\cite{Mingo-PRB-negf} for the nonlinear interactions.  Our approach is
essentially a generalization of the nonequilibrium Green's function
method in electronic transport \cite{meir-wingreen,haug}, with
fermions replaced by bosons as basic entities.  Although the
techniques have been extensively described in the literature, most of
them are centered around fermions.  In this paper, we give a
description of the method, emphasizing on both the theoretical
formulation and computational implementation.  We begin with an
elementary introduction of the Green's functions, including the
contour-ordered Green's functions.  We then discuss equations of
motion of the Green's functions, Feynman diagrammatic expansion, and
Dyson equations.  We discuss the heat current and derive a formula for
effective transmission when there are nonlinear interactions.  After
the introduction of the method, we present results of one-dimensional
(1D) chains, benzene ring, and carbon nanotubes, and discuss some of
the interesting features in such systems.

\section{Nonequilibrium Green's Function Method}

\subsection{Definition of Green's functions\label{sec-green}}
The nonequilibrium Green's function methods are discussed in
Refs.~\cite{kadnoff-baymn,haug}, and equilibrium Green's functions are
explained in many textbooks such as Refs.~\cite{doniach,mahan}.  For
completeness, we introduce our notations and give a quick review of
the definitions and properties of various Green's functions in this
subsection.  We define the retarded Green's function as
\begin{equation}
G^r(t,t') = - \frac{i}{\hbar}\theta(t-t') \langle [u(t), u(t')^T] \rangle,
\end{equation}
where the column vector $u$ with component $u_j$ is not the usual
displacement operator from equilibrium but a renormalized one, i.e.,
$u_j = x_j\sqrt{m_j}$, such that the kinetic energy is of the form
$(1/2) \dot{u}^T \dot{u}$.  The superscript $T$ stands for matrix
transpose.  The square brackets are the commutators.  $G^r$ is a
square matrix with elements $G^r_{jk}(t,t') = - (i/\hbar)
\theta(t-t')\langle [u_j(t), u_k(t')]\rangle$.  The physical dimension
of $G^r$ is time. By definition, $G^r(t,t')$ equals zero when $t
\leq t'$.  In equilibrium or nonequilibrium steady states, the Green's
function depends only on the difference in time, $t-t'$.  The Fourier
transform of $G^r(t-t') = G^r(t,t')$ is defined as
\begin{equation}
G^r[\omega] = \int_{-\infty}^{+\infty}\!\!\! G^r(t) e^{i\omega t}dt.
\end{equation}
Note that we use square brackets to delimit the argument for the
Green's functions in frequency domain.  The dimension of the retarded
Green's function in frequency domain is time squared.  The inverse
transform is given by
\begin{equation}
G^r(t) = \frac{1}{2\pi}\int_{-\infty}^{+\infty}\!\!\! G^r[\omega] e^{-i\omega t}d\omega.
\end{equation}
For notational simplicity, we'll set $\hbar = 1$.  We can always get
the final required expressions by a simple dimension analysis.

The rest of the definitions are the advanced Green's function
\begin{equation}
G^a(t,t') = i\theta(t'-t) \langle [u(t), u(t')^T] \rangle,
\end{equation}
the ``greater than'' Green's function
\begin{equation}
G^{>}(t,t') = - i \langle u(t) u(t')^T \rangle,
\end{equation}
the ``less than'' Green's function
\begin{equation}
G^{<}(t,t') = - i \langle u(t') u(t)^T \rangle^T,
\end{equation}
the time-ordered Green's function
\begin{equation}
G^{t}(t,t') = \theta(t-t') G^{>}(t,t') + 
              \theta(t'-t) G^{<}(t,t'),
\end{equation}
and anti-time-ordered Green's function
\begin{equation}
G^{\bar{t}}(t,t') = \theta(t'-t) G^{>}(t,t') + 
              \theta(t-t') G^{<}(t,t').
\end{equation}
The following linear relations hold both in frequency and time domains
from the basic definitions:
\begin{eqnarray}
G^r - G^a &=& G^{>} - G^{<},\\
G^t + G^{\bar{t}} &=& G^{>} + G^{<},\\
G^t - G^{\bar{t}} &=& G^r + G^a.
\end{eqnarray}
Also, the relations $G^r = G^t - G^{<}$ and $G^a = G^{<}-G^{\bar{t}}$
are useful.  Out of the six Green's functions, only three of them are
linearly independent.  However, in systems with time translational
invariance, the functions $G^r$ and $G^a$ are Hermitian conjugate of
one other:
\begin{equation}
G^a[\omega] = (G^r[\omega])^\dagger.
\end{equation}
So in general nonequilibrium steady-state situations, only two of them
are independent.  In this paper, we consider them to be $G^r$ and
$G^{<}$.  There are other relations in the frequency domain as well:
\begin{eqnarray}
G^{<}[\omega]^\dagger &\!=\!& - G^{<}[\omega],\\
G^r[-\omega] &\!=\!& G^r[\omega]^*,\\
G^{<}[-\omega] &\!=\!& G^{>}[\omega]^T \!=\! 
- G^{<}[\omega]^* \!\!+\! G^r[\omega]^T \!\!-\! G^r[\omega]^*.
\end{eqnarray}
The last two equations show that we only need to compute the positive
frequency part of the functions.

In thermal equilibrium, there is an additional equation relating $G^r$
and $G^{<}$:
\begin{equation}
G^{<}[\omega] = f(\omega) \Bigl( G^r[\omega] - G^a[\omega] \Bigr),
\label{GrtoGl}
\end{equation}
where 
\begin{equation}
f(\omega) = \frac{1}{ \exp(\beta \hbar \omega) - 1}
\end{equation}
is the Bose-Einstein distribution function at temperature $T=1/(k_B
\beta)$.  Equation~(\ref{GrtoGl}) is obtained by writing the Green's
functions as a sum of energy eigenstates (known as the Lehmann
representation).  Thus in equilibrium, there is only one independent
Green's function; we take it to be $G^r$.

The contour-ordered Green's function is a convenient book-keeping to
treat the different Green's functions in a concise notation.  We can
consider a contour-ordered Green's function as a function $G(\tau,
\tau')$ with arguments $\tau$ and $\tau'$ defined on the complex
plane.  The contour runs from $-\infty$ slightly above the real axis
to $+\infty$ and loops back from $+\infty$ slightly below the real
axis to $-\infty$.  The contour-ordered Green's function can be mapped
onto four different normal Green's functions by
$G^{\sigma\sigma'}(t,t') = \lim_{\epsilon \to 0^+} G(t\!  +\! i
\epsilon \sigma, t'\!+\! i\epsilon \sigma')$, where $\sigma = \pm
(1)$, and $G^{++} = G^{t}$ is the time-ordered Green's function,
$G^{--} = G^{\bar{t}}$ is the anti-time-ordered Green's function,
$G^{+-} = G^{<}$, and $G^{-+} = G^{>}$.

In a noninteracting harmonic system with a Hamiltonian
\begin{equation}
H_0 = \frac{1}{2} \dot{u}^T \dot{u} + \frac{1}{2} u^T K u,
\end{equation}
the retarded Green's function in the frequency domain is given by
\begin{equation}
G^r[\omega] = \left( (\omega + i \eta)^2 I - K \right)^{-1},
\label{Gr-free}
\end{equation}
where $I$ is an identity matrix, and $\eta \to 0^{+}$.  Adding a small
$\eta$ helps to choose correctly the inverse Fourier transform
integration path in the complex $\omega$-plane, so that the retarded
Green's function has the required causal property, $G^r(t) = 0$ for $t
< 0$.  Equation~(\ref{Gr-free}) can be derived by an equation of
motion method. We note that the retarded Green's function is a
symmetric matrix since $K$ is symmetric, but this feature is not
preserved with nonlinear interactions.

\subsection{Model and adiabatic switch-on}

Our system is a junction with a central region and two leads which
serve as heat reservoirs.  We treat the leads explicitly as
quasi-one-dimensional periodic lattices.  This setup is experimentally
relevant and conceptually useful for computation.  We consider
non-conducting solid and treat only the vibrational degrees of freedom
for heat transport.  Let the (mass-normalized) displacement from some
equilibrium position for the $j$th degree of freedom in the region
$\alpha$ be $u_j^\alpha$; $\alpha = L, C, R$, for the left, center,
and right regions, respectively. The quantum Hamiltonian is given by
\begin{equation}
{\cal H} = \!\!\!\!\!\sum_{\alpha=L,C,R}\!\!\!\!\!H_\alpha  + (u^L)^T V^{LC} u^C + (u^C)^TV^{CR} u^R + H_n,
\end{equation}
where $H_{\alpha} = \frac{1}{2} {(\dot{u}^\alpha)}^T \dot{u}^\alpha +
\frac{1}{2} {(u^\alpha)}^T K^\alpha u^\alpha$, $u^\alpha$ is a column
vector consisting of all the displacement variables in region
$\alpha$, and $\dot{u}^\alpha$ is the corresponding conjugate
momentum.  $K^\alpha$ is the spring constant matrix and
$V^{LC}=(V^{CL})^T$ is the coupling matrix of the left lead to the
central region; similarly for $V^{CR}$.  We note that the dynamic
matrix of the full linear system is
\begin{equation}
K = \left( \begin{array}{lll} K^L & V^{LC} & 0 \\
             V^{CL} & K^C & V^{CR} \\
             0   & V^{RC} & K^R
           \end{array}\right).
\end{equation}
The nonlinear part of the interaction will take the form
\begin{equation}
H_n = \frac{1}{3} \sum_{ijk} T_{ijk}\, u_i^C u_j^C u_k^C + 
\frac{1}{4} \sum_{ijkl} T_{ijkl}\, u_i^C u_j^C u_k^C u_l^C.
\end{equation}
The quartic interaction is important in stablizing the system, as a
purely cubic nonlinear interaction makes the energy unbounded from
below.

\begin{figure}
\includegraphics[width=\columnwidth]{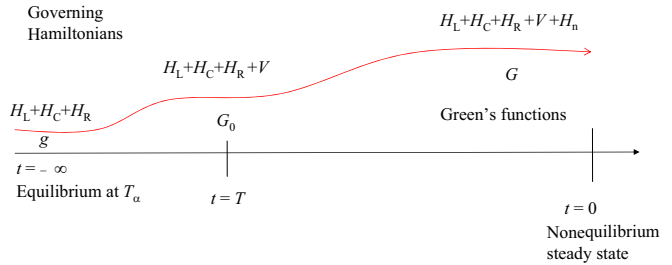}
\caption{\label{fig-adiabatic}A schematic to illustrate the two
adiabatic switch-ons.}
\end{figure}

We need to answer an important question: what is the distribution of
the system?  The distribution enters the definition of the Green's
functions as a density matrix $\rho(0)$ for the average $\langle
\cdots \rangle$.  For an equilibrium problem, this is just the
Boltzmann factor, but in a nonequilibrium situation, it is not known
and must be computed in some way.  The adiabatic switch-on gives us a
clear conceptual framework how this problem can be solved, at least
formally.  We imagine that at $t = -\infty$ the system has three
decoupled regions, each at separate temperatures, $T_L$, $T_C$, and
$T_R$.  The nonlinear interactions are turned off.  The Green's
functions $g^\alpha$ are known and take the form of
Eq.~(\ref{Gr-free}).  The couplings $V^{LC}$ and $V^{CR}$ are then
turned on slowly, and a steady state of the linear system is
established at some time $T \ll 0$.  The Green's functions of the
linear nonequilibrium system will be denoted by $G_0$.  For this
linear problem, the result does not depend on $T_C$.  Finally, the
nonlinear interaction $H_n$ is turned on, and at time $t=0$, a
nonequilibrium steady state is established, see
Fig.~\ref{fig-adiabatic} for illustration.  The full nonlinear Green's
functions will be denoted by $G$.

The density matrices at time $t=-\infty$, $T$, and $t=0$ are related
in the following way in the interaction picture:
\begin{eqnarray}
\rho(T) &=&  S_0(T,-\infty) \rho(-\infty) S_0(-\infty, T),\\
S_0(t,t') &=& {\cal T} e^{-i \int_{t'}^{t} V(t'')dt''},\\
\rho(0) &=& S(0,T) \rho(T) S(T,0),\\
S(t,t') &=& {\cal T} e^{-i \int_{t'}^{t} H_{n}(t'')dt''},
\end{eqnarray}
where $\cal T$ is the time-order operator. 

\subsection{Equation of motion method}
In this subsection, we derive the equations of motion for Green's
functions.  The equations of motion can be used to perform systematic
perturbation expansion for the Green's functions, or as a starting
point for mean-field approximations.  There are at least two ways to
derive the equations of motion for nonequilibrium Green's functions.
The first is to derive the equations of motion for the time-ordered
Green's function and then generalize it to the contour-ordered version
by evoking the structure isomorphism of the two sets of Green's
functions \cite{meir-wingreen}.  Another possibility is to consider
directly the contour-ordered Green's function \cite{niu}.  We'll take
the latter approach.

We define a general $n$-point contour-ordered Green's function as
\begin{eqnarray}
G_{j_{1}, j_{2}, \cdots, j_{n}}^{\alpha_{1}, \alpha_{2}, \cdots, \alpha_{n}}
(\tau_1, \tau_2, \cdots, \tau_n) = \qquad\qquad\qquad\qquad \nonumber \\
\qquad\qquad\qquad - i \langle {\cal T}_{\tau} u_{j_1}^{\alpha_1}(\tau_1) 
u_{j_2}^{\alpha_2}(\tau_2) \dots 
u_{j_n}^{\alpha_n}(\tau_n)\rangle.
\end{eqnarray}
The index $\alpha =L, C, R$ labels the region, $j$ labels the degrees
of freedom in that region, and $\tau$ is the contour variable.  The
function is symmetric with respect to simultaneous permutations of the
triplet $(\alpha, j, \tau)$.  We view the function as an analytic
function (at least along the path) in variable $\tau$ which varies on
the Keldysh contour \cite{keldysh}.  We also introduce a generalized
$\theta(\tau,\tau')$ function which is defined to be 1 if $\tau$ is
later than $\tau'$ along the contour and 0 otherwise.  Its derivative
is a generalized $\delta$-function, $\delta(\tau-\tau') = \sigma
\delta_{\sigma, \sigma'} \delta(t-t')$.  The $\theta$ function can be
used to represent the contour-order operator ${\cal T}_{\tau}$.  For
convenience, we allow for $n=0$, which is just a constant $-i$.  Our
definition is slightly different from that of the usual quantum field
theory where our prefactor is always $-i$ (instead of the usual
$(-i)^{n/2}$).

To start with, we consider one of the important correlation functions
for heat transport, $G^{C,L}$.  The first derivative with respect to
the second argument always leads to a direct differentiation inside
the contour-order operator since the equal-time coordinates $u$
commute,
\begin{equation}
{\partial G_{j,l}^{C,L}(\tau, \tau') \over \partial \tau' } 
= - i \langle {\cal T}_{\tau} u_j^C(\tau) \dot{u}_l^L(\tau') \rangle.
\end{equation}
The second derivative is similar in this case as $u^C$ and $\dot{u}^L$
also commute at equal time. Thus, substituting the equation of motion
for $u_L$ (having identical form in quantum and classical cases), we
get
\begin{equation}
{\partial^2 G_{j,l}^{C,L}(\tau,\! \tau') \over \partial \tau'^2} 
= \!-\!\! \sum_{m} G_{j,m}^{C,L}(\tau,\! \tau') K^L_{ml}
-\!\! \sum_{m} G_{j,m}^{C,C}(\tau,\! \tau') V_{ml}^{CL}.
\label{eqG_CL-component}
\end{equation}
In matrix notation, it is 
\begin{equation}
{\partial^2 G^{C,L}(\tau, \tau') \over \partial \tau'^2} 
+ G^{C,L}(\tau, \tau') K^L = 
-  G^{C,C}(\tau, \tau') V^{CL}. \label{eqG_CL}
\end{equation}
To solve this differential equation on the contour, we define the
(contour-ordered) Green's function of the left lead to satisfy
\begin{equation}
{\partial^2 g^L(\tau', \tau'') \over \partial \tau'^2} 
+ K^L g^L(\tau', \tau'') = -I \delta(\tau'-\tau''). \label{eqg_L}
\end{equation}
We can think of $-g^L$ as the inverse of the operator
$\partial^2/\partial \tau'^2 + K^L$.  By multiplying
$g^L(\tau',\tau'')$ to Eq.~(\ref{eqG_CL}) from the right,
$G^{C,L}(\tau, \tau') $ to Eq.~(\ref{eqg_L}) from the left, and then
subtracting the two equations and integrating over the variable
$\tau'$ along the contour, we get (after integration by part)
\begin{eqnarray}
G^{C,L}(\tau,\tau'') = \int G^{C,C}(\tau,\tau') V^{CL} g^L(\tau', \tau'') d\tau' + 
\qquad \nonumber \\
{\partial G^{C,L}\!(\tau,\! \tau') \over \partial \tau' } g^L\!(\tau',\!\tau'') 
\!+\! G^{C,L}\!(\tau, \!\tau'){\partial g^L\!(\tau',\!\tau'') 
\over \partial \tau'}\Big|^{\tau' = -\infty - i\epsilon}_{\tau' = -\infty +i\epsilon}. 
\end{eqnarray}
The reason that the terms in the second line are identically zero
involves some considerations.  First, since the central part and left
lead are decoupled at the time $\tau = -\infty +i \epsilon$, we must
have $G^{C,L}(\tau, -\infty + i \epsilon) = 0$ and ${\partial\over
\partial \tau' } G^{C,L}(\tau, -\infty + i \epsilon) = 0$.  This takes
care of the lower bound.  Next, we require that the free left lead
Green's function satisfies the boundary condition $g^L(\tau',\tau'') =
0$, ${\partial \over \partial \tau' } g^L(\tau',\tau'') = 0$, at the
upper limit $\tau' = - \infty - i\epsilon$.  This requirement is
equivalent to $g^{\bar t}(-\infty) = 0$, and $g^{>}(-\infty) = 0$, or
$g^r(-\infty) = 0$ (as well as that their first derivatives are zero).
The condition $g^{r}(-\infty) = 0$ is consistent with the causality
requirement of the retarded Green's functions, but the conditions for
$g^{\bar {t}}$ or $g^{>}$ are insufficient, as the information about
temperature of the system is unspecified.  We fix the ambiguity by
requiring that
\begin{equation}
 g^{>}(t=0) = - i \langle u^L(0) u^L(0)^T \rangle
\end{equation}
be the equilibrium value at inverse temperature $\beta_L$.  In other
words, $g^L$ should exactly be the contour-ordered Green's function of
the free lead as defined in subsec.~\ref{sec-green}.  Such a choice is
consistent with the adiabatic switch-on.  A full justification of the
final result,
\begin{equation}
G^{C,L}(\tau,\tau'') = \int G^{C,C}(\tau,\tau') V^{CL} g^L(\tau', \tau'')\, d\tau',
\label{GCL}
\end{equation}
can also be given from a perturbation expansion, as has been done in
Ref.~\cite{meir-wingreen}.

Equation~(\ref{GCL}) can be generalized to the case of $G^{C,C,L}$
with a similar result.  We now consider the Green's functions of the
form $G^{C,C,\cdots,C}$ involving only the central part.  The aim is
to derive a recursion relation for the central part Green's functions
by eliminating references to full Green's functions involving the
leads.  This type of Green's functions involves two new features, (1)
we must take care of the derivatives of the $\theta$ functions which
produce commutators $[u^{C}, \dot{u}^{C}]$, (2) we generate higher
order Green's functions due to the nonlinear interactions.

The contour order leads to $n!$ terms, each of which is a permutation
of the displacement variable $u$ from the original order
$1,2,\cdots,n$, multiplied by a term of permutation of the arguments
$\tau$ from the standard order $\theta(\tau_1 - \tau_2) \theta(\tau_2
- \tau_3) \cdots \theta(\tau_{n-1} - \tau_n)$.  Differentiation of
these $\theta$ functions leads to a sum of contractions between the
variable that we are differentiating and all the other variables.
This reduces the order of the Green's function by two.  We note that
\begin{equation}
[ u_j^C(\tau),  \dot{u}_l^C(\tau) ]  = i \delta_{jl}.
\end{equation}

For the moment, we'll consider only the cubic nonlinearity.  Four
terms are produced when $\ddot{u}^C$ is substituted inside the order
operator, due to linear coupling $K^C$, $V^{CL}$, $V^{CR}$, and
nonlinear $T_{ijk}$.  The first three terms are similar in structure
to that in Eq.~(\ref{eqG_CL-component}).  The cubic nonlinear
interaction increases the value of $n$ by 1.  We note that the
equation for $\ddot{u}^C$ is evaluated at the same time $\tau$.  This
would be complicated if we have to keep track of which $\tau$'s are
equal in the Green's functions.  We transfer this information to the
coupling $T_{ijk}$ to define the three-body interaction in contour
variable form (for the force)
\begin{equation}
- \sum_{jk} \int\int T_{ijk}(\tau, \tau', \tau'') u_{j}^C(\tau')u_{k}^C(\tau'')\, d\tau' d\tau''.
\end{equation}
Since the contour integral is $\int d\tau = \sum_{\sigma}
\sigma\int_{-\infty}^{+\infty} dt$, we must define
\begin{equation}
T_{ijk}(\tau,\tau',\tau'') = T_{ijk}
\sigma' \delta_{\sigma,\sigma'} \delta(t-t') 
\sigma'' \delta_{\sigma,\sigma''} \delta(t-t''). 
\end{equation}
This is because our $\delta$-function has the standard property that
$\int \delta(\tau - \tau') d\tau = 1$, but when written in the
ordinary time $t$ and index $\sigma$, we must have the extra $\sigma$
factor to get the required value of $+1$.

\begin{figure}
\includegraphics[width=\columnwidth]{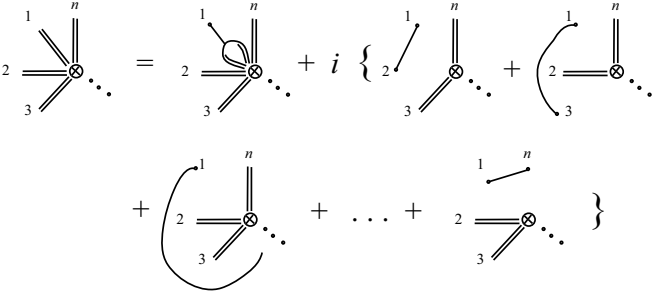}
\caption{\label{fig-exprule}Recursive expansion rule for Green's
functions. A vertex with $n$ double lines denotes an $n$-point Green's
function.  The single line denotes $G_0$. A single-line three-terminal
vertex is associated with $T(\tau,\tau',\tau'')$.}
\end{figure}

The two terms involving the leads produce Green's functions of the
form $G^{C,C,...,L}$ and $G^{C,C,...,R}$, which can be replaced by
$G^{C,C,...,C}$ using the result Eq.~(\ref{GCL}) derived earlier.  We
can now solve for the center-only Green's function using the Green's
function of the linear system:
\begin{eqnarray}
\left(\frac{\partial^2}{\partial \tau^2} + K^C\right) G_0(\tau,\tau') + 
\int d\tau'' \Big( V^{CL}g^L(\tau,\tau'')V^{LC} +
\nonumber \\
\label{eq_G0}
V^{CR}g^R(\tau,\tau'')V^{RC}\Big)G_0(\tau'',\tau') = -I \delta(\tau-\tau').
\quad
\end{eqnarray}
We then obtain the following simple recursive rules for Green's
functions (as shown in Fig.~\ref{fig-exprule}):

(1) Replace leg 1 by inserting a nonlinear coupling $T(\tau, \tau',
\tau'')$ such that the outer leg is immediately connected with $G_0$
while the other two terminals increase the order of the Green's
function by 1.  Quartic interaction is similar, but the process will
increase the order by 2.

(2) Add imaginary unit $i$ times a sum of the $n-1$ graphs formed by pairing
each leg with leg 1 and connecting with the propagator $G_0$,
multiplied by a $(n-2)$ order remaining Green's function.

(3) Symmetrize the graphs, if desired, i.e., do steps (1) and (2) for
every leg, 1, 2, $\cdots$, $n$, add them up, and then divide by $n$.

The above rules may be conveniently implemented in a symbolic computer
language, such as Mathematica.

\subsection{Feynman diagrams and Dyson equations}
The contour-ordered Green's function can also be obtained by a
perturbation expansion of the interaction picture evolution operator
(scattering matrix operator) and is expressed as:
\begin{eqnarray}
G_{jk}(\tau, \tau') & = & - i \langle {\cal T}_{\tau} u_j^H(\tau) u_k^H(\tau') \rangle
\nonumber\\
\label{expand} & = & - i \langle {\cal T}_{\tau} u_j^I(\tau) u_k^I(\tau') 
e^{-i \int H_n^I(\tau'')d\tau''}  \rangle_0,\quad
\end{eqnarray}
where the displacements refer to the central region; we have dropped
the superscript $C$ for brevity.  The operators in the top line are in
the Heisenberg picture, while that in the second line are in the
interaction picture.  The Green's function $G_0$ of the linear system
(when $H_n=0$) can be computed from that of the free subsystems (an
integral equation form of Eq.~(\ref{eq_G0})):
\begin{equation}
G_0(\tau,\!\tau')\! = \! g^C(\tau,\!\tau')\! +\! 
\int\!\! d\tau_1 d\tau_2\, g^C(\tau,\!\tau_1) \Sigma(\tau_1,\!\tau_2) G_0(\tau_2,\!\tau'),
\label{dyson0}
\end{equation}
where $\Sigma = \Sigma_L + \Sigma_R$, 
\begin{equation}
\Sigma_L = V^{CL} g^L V^{LC},
\end{equation}
similarly for $\Sigma_R$.  This Dyson equation can be derived by
considering the first step of the adiabatic switching-on process.
Since it is a linear system, the self-energy $\Sigma$ is known
exactly.

By expanding the exponential in Eq.~(\ref{expand}), a series in the
nonlinear interaction strength is obtained.  The reduction in terms of
the unperturbed Green's function $G_0$ relies on the fact that Wick's
theorem \cite{fetta-walacka} is applicable here.  Although a formal
proof of Wick's theorem is difficult, we can see that the density
matrix $\rho(0)$ must be quadratic in exponential.  This is because
the $\rho(-\infty)$ is quadratic and the system is linear.  The fact
that the equation of motion method and the perturbation expansion give
identical results is a confirmation of the validity of Wick's theorem.

The expansion contains connected as well as disconnected Feynman
diagrams.  The disconnected diagrams are constant in time, and give
rise to a thermal expansion effect.  We can show that these diagrams
do not contribute to the thermal transport, as they are proportional
to $\delta(\omega)$ in the frequency domain.  The thermal current
formula has a factor of $\omega$ which makes it zero.

Finally, the connected part of the Green's function satisfies a
similar contour-ordered Dyson equation relating $G_c$ to $G_0$ through
a nonlinear self-energy $\Sigma_n$:
\begin{equation}
G_c(\tau,\!\tau')\! = \! G_0(\tau,\!\tau')\! +\! \!
\int\!\! d\tau_1 d\tau_2 G_0(\tau,\!\tau_1) \Sigma_n(\tau_1,\!\tau_2) G_c(\tau_2,\!\tau').
\label{dyson}
\end{equation}
In ordinary Green's functions and in frequency domain ($\omega$
argument suppressed), the Dyson equations have solutions \cite{haug}:
\begin{eqnarray}
G_0^r &\!=\!& {G_0^a}^\dagger = \left( (\omega+i\eta)^2 I \!-\! K^C 
\!-\! \Sigma^r \right)^{-1}, \label{G1} \\
G_0^{<} &\!=\!& G_0^r \Sigma^{<} G_0^a, \\
G_c^r &\!=\!& \left( {G_0^r}^{-1} - \Sigma^r_n \right)^{-1}, \\
G_c^{<} &\!=\!& G_c^r \bigl(\Sigma^{<} + \Sigma^{<}_n\bigr) G_c^a .
\label{G4}
\end{eqnarray}
The connected part of the Green's function $G_c$ will be used for $G$
in Eq.~(\ref{heat-current}) below, and the extra subscript $c$ will be
dropped from this point.

\begin{figure}
\includegraphics[width=\columnwidth]{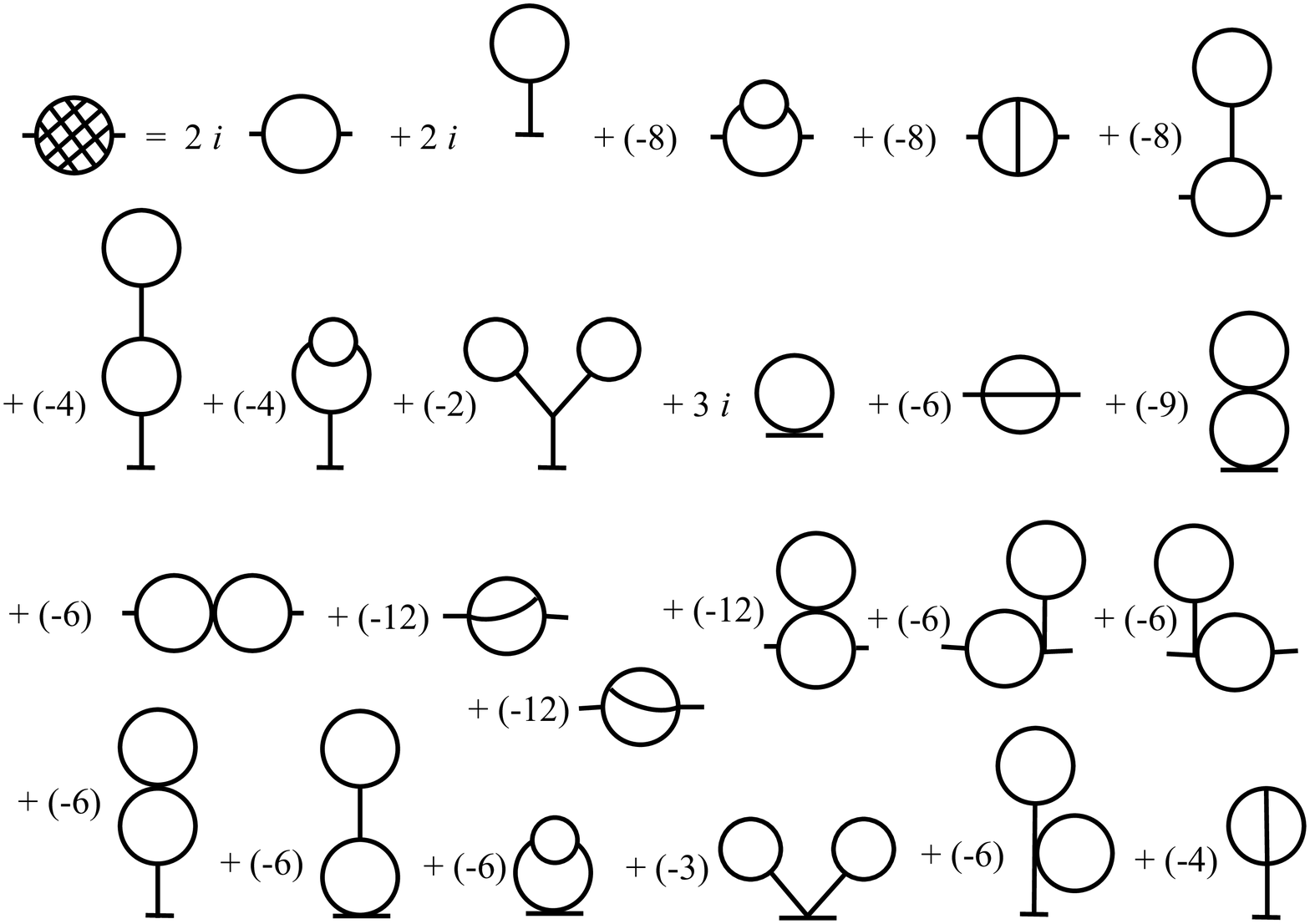}
\caption{\label{fig-feynmant4}The Feynman diagrams for nonlinear
self-energy $\Sigma_n$ with the prefactors for graphs of cubic and
quartic interactions to second order in $\hbar$.}
\end{figure}

In Fig.~\ref{fig-feynmant4} we give the Feynman diagrams to ``second
order'' of the nonlinear interactions for cubic and quartic terms,
where the graphs have explicitly $\hbar$ and $\hbar^2$ in the
coefficients; the neglected graphs have higher powers of $\hbar$.
These diagrams are the same as those in Ref.~\cite{valle}, but the
interpretation is different.  It is for the contour-ordered Green's
functions here and for the retarded Green's functions of the
creation/annihilation operators in Ref.~\cite{valle}.  To find the
actual expressions for each of the diagrams, we have the following
simple rules: (1) label each interaction vetex line by the $\tau$
variables and site indices, with the interaction function
$T_{ijk}(\tau_i, \tau_j, \tau_k)$ (or $T_{ijkl}(\tau_i, \tau_j,
\tau_k, \tau_l)$ for the quartic interaction).  (2) Each line
connecting the labels is associated with a propagator $G_0$.  (3)
Integrate over all the internal variables along the contour; sum over
the internal site indices.  Most of the integrations are easy, as
$T_{ijk}(\cdots)$ contains two $\delta$-functions in $\tau$.  For
example, the second graph is
\begin{eqnarray}
\sum_{j_3,j_4,j_5,j_6}\!\! \int d\tau_3 d\tau_4 d\tau_5 d\tau_6\, T_{j_1, j_2, j_3}(\tau_1, \tau_2, \tau_3)
\qquad\qquad
\nonumber\\
\qquad G_{0,j_3j_4}(\tau_3, \tau_4) 
T_{j_4,j_5,j_6}(\tau_4,\tau_5,\tau_6) G_{0,j_5j_6}(\tau_5, \tau_6),
\end{eqnarray}
where $\tau_1$ and $\tau_2$ are external variables and the rest are
internal.  After integrations, we get
\begin{equation}
\!\!\!\!\!\!\!\!\sum_{j_3,j_4,j_5,j_6}\!\!\!\!\!\!\!\int\! d\tau_4  T_{j_1, j_2, j_3}\! \delta(\tau_1- \tau_2) G_{0,j_3j_4}\!(\tau_1,\! \tau_4) 
T_{j_4,j_5,j_6}\! G_{0,j_5j_6}(\tau_4,\! \tau_4).
\end{equation}
Now we can rewrite in real time $t$ and the branch index $\sigma$
using the mapping $G(\tau, \tau') \to G^{\sigma\sigma'}(t, t')$, $\int
d\tau \to \sum_{\sigma=\pm 1} \int_{-\infty}^{\infty} \sigma dt$, and
$\delta(\tau -\tau') \to \sigma \delta_{\sigma, \sigma'}
\delta(t-t')$.  Noting that $G_0(\tau_4, \tau_4) = G_0(0)$ by time
translational invariance, we have
\begin{eqnarray}
\sigma_1 \delta_{\sigma_1, \sigma_2} \delta(t_1-t_2) \!\!\!\!\sum_{j_3,j_4,j_5,j_6,\sigma_4}\!\!T_{j_1,j_2,j_3} 
T_{j_4,j_5,j_6}\qquad\qquad\quad \nonumber \\
\qquad\qquad\qquad\quad \int G_{0,j_3,j_4}^{\sigma_1, \sigma_4}(t_1-t_4) 
G_{0,j_5,j_6}^{\sigma_4, \sigma_4}(0) \sigma_4\, dt_4.\qquad 
\end{eqnarray}
The result can now be transformed to frequency domain if desired.

\subsection{Thermal current and conductance}
We define the current flow from the left lead to the central region as
\begin{equation}
I_L = - \langle \dot{H}_L(t) \rangle.
\end{equation}
The interpretation of the current is somewhat subtle as the leads are
semi-infinite in extent.  The definition is meaningful.  By the
Heisenberg equation of motion, we obtain, at $t=0$, $I_L = \langle
(\dot{u}^L)^T V^{LC} u^C \rangle$.  The expectation value can be
expressed in terms of a Green's function $G^{<}_{CL}(t,t') = - i
\langle u^L(t') u^C(t)^T \rangle^T$. Using the fact that operators $u$
and $\dot{u}$ are related in Fourier space as $\dot{u}[\omega] =
(-i\omega) u[\omega]$, we can eliminate the derivative and get,
\begin{equation}
I_L = - \frac{1}{2\pi} \int_{-\infty}^\infty \!\!\!\!{\rm Tr}\left(
V^{LC} G^{<}_{CL}[\omega]\right) \omega\, d\omega.
\end{equation}
Using the result derived earlier relating the mixed lead-center
Green's function to the center-only Green's function and applying the
Langreth theorem \cite{langreth,haug} to write contour-ordered Green's
function in ordinary Green's functions in frequency domain, we have
$G^{<}_{CL}[\omega] = G^r_{CC}[\omega] V^{CL} g^{<}_L[\omega]
+G^{<}_{CC}[\omega]V^{CL}g^{a}_L[\omega]$.  The final expression for
the energy current is
\begin{equation}
I_L = - \frac{1}{2\pi}\!\! \int_{-\infty}^{+\infty}\!\!\!\!\!\!\!\! d\omega\, \omega
\, {\rm Tr}\Bigl( G^r[\omega] \Sigma^{<}_L[\omega] + 
G^{<}[\omega] \Sigma^a_L[\omega] \Bigr),
\label{heat-current}
\end{equation}
where the self-energy due to the interaction with the lead is
$\Sigma_L = V^{CL} g_{L} V^{LC}$.  For simplicity, we have dropped the
subscript $C$ on the Green's functions denoting the central
region. This result is the phonon analog of the electronic current
formula \cite{meir-wingreen,haug}.

Since the energy current must be conserved, we have $I_L = - I_R$.  In
addition, $I_L$ is real. We can obtain a symmetrized expression for
the current
\begin{eqnarray}
I = \frac{1}{4}(I_L + I_L^{*} - I_R - I_R^{*}) =
\frac{1}{4\pi} \int_{0}^{\infty} d\omega \qquad\qquad\nonumber \\
\omega\, 
{\rm Tr}\Bigl\{(G^r-G^a)(\Sigma_R^{<}-\Sigma_L^{<}) + i G^{<}
(\Gamma_R-\Gamma_L) \Bigr\}, \label{current-2}
\end{eqnarray}
where 
$\Gamma_\alpha = i(\Sigma_\alpha^r - \Sigma_\alpha^a)$.

We define the thermal conductance as
\begin{equation}
\bar{\kappa} = \lim_{\Delta T \to 0} \frac{I}{\Delta T} = 
\frac{\delta I}{\delta T},
\label{conductance-def}
\end{equation}
where $\Delta T$ is the difference of the temperatures in leads, such
that $T_L = T + \Delta T/2$ and $T_R = T - \Delta T/2$.  Using
Eq.~(\ref{current-2}), we can actually take the limit $\Delta T \to 0$
by introducing variational derivatives of the Green's functions,
self-energies, or current as, e.g.,
\begin{equation}
 \frac{\delta G}{\delta T} = \lim_{\Delta T \to 0 }
\frac{G(T_L, T_R) - G(T, T)}{T_L - T_R}.
\end{equation}
The difference in the numerator is the value of a function when the
leads are at two different temperatures $T_L$ and $T_R$, respectively,
minus the value when the system is in equilibrium at $T_L = T_R = T$.
With this notation, we can express the conductance in a form similar
to the Landauer formula \cite{landauer} for the ballistic transport as
\begin{equation}
\label{conductance-2}
\bar{\kappa} = 
\frac{1}{2\pi} \int_0^\infty \!\!\!d\omega\, \omega\, \tilde{T}[\omega] 
\frac{\partial f(\omega)}{\partial T}
\end{equation}
with an effective transmission coefficient,
\begin{eqnarray}
\tilde{T}[\omega] &=& 
\frac{1}{2} {\rm Tr} \bigl\{ G^r (\Gamma_L + \frac{1}{2}\Gamma_n - S) G^a \Gamma_R \bigr\} +\nonumber \\
\label{eff-trans}
&&\frac{1}{2} {\rm Tr} \bigl\{ G^a \Gamma_L G^r (\Gamma_R + \frac{1}{2}\Gamma_n +S) \bigr\},
\end{eqnarray}
where the nonlinear effect is reflected in the extra terms,
$\Gamma_n = i (\Sigma_n^r - \Sigma_n^a)$, and 
\begin{equation}
S = i\left[ f \bigl(\frac{\delta \Sigma_n^r}{\delta T} - 
\frac{\delta \Sigma_n^a}{\delta T}\bigr)  - 
\frac{\delta \Sigma_n^{<}}{\delta T} \right]
\left({\partial f \over\partial T}\right)^{-1}.
\end{equation}
Equation~(\ref{eff-trans}) is a generalization of the Caroli formula
\cite{caroli} for ballistic transport.  We'll discuss in
Sec.~\ref{sec3b} how the variational derivatives of the self-energy
can be computed.

\begin{figure}
\includegraphics[width=\columnwidth]{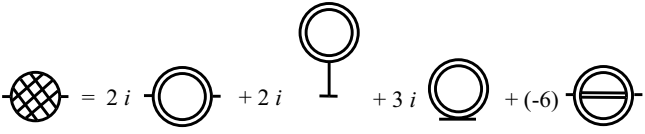}
\caption{\label{fig-mf}Mean-field approximation to the self-energy.
The double line denotes the full Green's function $G$; the single line
is for $G_0$.}
\end{figure}

\subsection{Mean-field approximation}
We have seen in Ref.~\cite{PRB-green} that high-order Feynman diagrams
are important at high temperatures.  However, it is computationally
very expensive to include these high-order diagrams, although we can
do calculations on very small systems such as a three-site model.  The
standard technique is to use a mean-field approximation which in some
sense partially takes into account these high-order graphs.  We
propose an approximation to the self-energy shown in
Fig.~\ref{fig-mf}, where the double line represents the nonlinear full
Green's function $G$, while a single line is the usual bare Green's
function $G_0$.  With such a choice, we find that to next order, all
the cubic interaction graphs are reproduced exactly (including the
prefactor), except that two of the diagrams are missing.

With the self-consistent mean-field approximation, the solution has to
be found iteratively.  The convergence of the iterative process
becomes an issue.  With only cubic interactions, the system of
equations will not converge at high temperatures.  We understand that
this is simply a consequence of the unstable nature when only cubic
terms are used.  We thus add the fourth order potential which is
essential for stability.  However, the convergence is still difficult
at high temperatures, mainly due to the singular nature of the Bose
distribution.

\subsection{Classical limit, molecular dynamics}
At high temperatures, the classical molecular dynamics (MD) is a
correct and numerically exact method for the heat conduction problem.
However, the standard heat baths used in MD such as Nos\'e-Hoover do
not correspond to exactly the action of the leads in our models.  A
correct heat bath should follow a Langevin dynamics with colored
noises.  Dhar \cite{Dhar-heat-bath} has given a derivation for the
junction problem.  The idea is to solve the lead degrees of freedom
$u^L$ and $u^R$ formally first, which is easy since it is a linear
system.  The result is then substituted into the equation for the
center region.  This gives the following equation:
\begin{equation}
\ddot{u}^C = -K^C u^C + F_n(u^C\!) \!- \!\!\int_{T}^t\!\! \Sigma^r(t, t')u^C(t') dt' 
+ \xi_L + \xi_R,
\label{langevin}
\end{equation}
where $F_n$ is the nonlinear force, $\Sigma^r$ is the same retarded
self-energy of the lead, $\Sigma^r = \Sigma_L^r +\Sigma_R^r$, used in
the nonequilibrium Green's function calculation, but in the time
domain.  By integration by part, we could also cast
Eq.~(\ref{langevin}) in a standard generalized Langevin form
\cite{G-Langevin} involving a damping force, but $K^C$ has to be
shifted to $K^C \!+\!  \Sigma^r[\omega\! =\! 0]$.  An extra
contribution from the left lead is
\begin{equation}
\xi_L(t) = V^{CL} \left( g^r_L(t,T) \dot{u}^L(T) 
- \frac{\partial g^r_L(t, T)}{\partial T} u^L(T) \right),
\label{random-noise}
\end{equation}
where $g^r_L$ is the retarded Green's function of the left lead in the
time domain.  The expression for the right lead $\xi_R$ is similar.  A
time $T$ in the lower limit in the integral and
Eq.~(\ref{random-noise}) is some time in the remote past where the
system is decoupled and the leads are in respective thermal
equilibrium.  We'll take the limit $T \to -\infty$.  We turn
Eq.~(\ref{langevin}) into a stochastic differential equation by
requiring that $u^L(T)$ and $\dot{u}^L(T)$ are random variables
satisfying the Boltzmann-Gibbs distribution
\begin{equation}
P(u^L, \dot{u}^L) \propto e^{-\beta_L \left( \frac{1}{2} (\dot{u}^L)^T 
\dot{u}^L + \frac{1}{2} (u^L)^T K^L u^L \right)}.
\end{equation}
Thus, $\xi_L$ is colored noise with the properties
\begin{eqnarray}
\langle \xi_L(t) \rangle & = & 0, \\
\beta_L \langle \xi_L(t) \xi_L(t')^T \rangle & = &
V^{CL} \Bigl( \dot{g}_L^r(t,T) \frac{1}{K^L}\, \dot{g}_L^r(t',T)^T \nonumber\\
  & &\!\!\! + \, g_L^r(t,T)\, g_L^r(t',T)^T \Bigr) V^{LC}\\
&&\!\!\!\!\!\!\!\!\!\!\!\!\!\!\!\!\!\!\!\! 
\,=\, - \int_{t}^\infty\!\! \Sigma_L^r(t'',t')\,dt'', \quad (t > t').
\end{eqnarray}
The last line of the above equation is a kind of
``fluctuation-dissipation'' theorem.  We note that the dot in $g$
denotes the derivative with respect to $T$.  For a sensible heat bath,
the noise should not depend on $T$, but only on the difference $t-t'$.
Indeed, this can be done if we rewrite the result in vibrational
eigen-modes.  We can write the Fourier transform of the noise
correlation as
\begin{equation}
{\rm Im}\; 2\, V^{CL} \left( (K^L)^{-1/2} 
\frac{1}{ (\omega^2  - i \eta)I - K^L} \right) V^{LC},
\end{equation}
where the square root of a matrix is defined by its eigenvalue
decomposition taking only the positive eigen frequencies.  This is
well defined since $K^L$ is positive definite.  The properties of the
right lead noise $\xi_R$ are analogous.  Such a set of stochastic
differential equations can be simulated on a computer.

\section{Implementation Details}

\subsection{Surface Green's functions}
The Green's functions of the free leads $g^r$ are required in
calculation for the lead self energies $\Sigma$.  Although the lead
Green's function is a semi-infinite matrix, due to a localized
coupling $V^{LC}$ or $V^{CR}$, only a corner set of elements is
required.  These submatrix elements are termed surface Green's
functions.  The surface Green's function can be computed rather
quickly by recursive iterations \cite{surface-green} or decimation.
For completeness, we give an algorithm in pseudo-code form.  The
reader is referred to the literature for the derivation of such
algorithm.

The program takes three square matrices of equal size for $k_{00}$,
$k_{11}$, and $k_{01}$.  The matrix $k_{00}$ is the block immediately
adjacent to the center, while $k_{11}$, and $k_{01} = k_{10}^{T}$ are
repeated for the semi-infinite chain of the lead.  They form the whole
dynamic matrix of the lead in the following way:
\begin{equation}
K^R = \left(\begin{array}{cccc} k_{00} & k_{01} & 0  & \cdots \\
                  k_{10} & k_{11} & k_{01} & 0 \\
                  0      & k_{10} & k_{11} & k_{01} \\
                  \cdots & 0      & k_{10} & \ddots 
           \end{array}
            \right).
\end{equation}
The left arrow `$\leftarrow$' below denotes assignment; $\epsilon$ is
an error tolerance.

\begin{verse}
$s \leftarrow k_{00}$\\
$e \leftarrow k_{11}$\\
$\alpha \leftarrow k_{01}$\\
$\beta \leftarrow \alpha^{T}$\\
$ g \leftarrow [(\omega+ i\eta)^{2} I - e]^{-1}$\\

do\\

\ \ \ \ $s' \leftarrow s + \alpha g \beta$\\
\ \ \ \ if $|s' -s| < \epsilon$ exit \\
\ \ \ \ $e' \leftarrow e + \alpha g \beta + \beta g \alpha$\\
\ \ \ \ $\alpha' \leftarrow \alpha g \alpha$ \\
\ \ \ \ $\beta' \leftarrow \beta g \beta$ \\
\ \ \ \ $s \leftarrow s'$ \\
\ \ \ \ $e \leftarrow e'$ \\
\ \ \ \ $\alpha \leftarrow \alpha'$ \\
\ \ \ \ $\beta \leftarrow \beta'$ \\
\ \ \ \ $ g \leftarrow [(\omega+ i\eta)^{2} I - e]^{-1}$\\

end do\\

\indent $ g^r \leftarrow [(\omega+ i\eta)^{2} I - s]^{-1}$.\\
\end{verse}

\subsection{\label{sec3b}Fast Fourier transform and variational derivatives}

The diagrams in the mean-field approximation can be computed in the
time domain easily and efficiently; the convolutions in frequency
domain become simple multiplications in time domain.  Thus we have
adapted a fast Fourier transform method for the calculation.  This
reduces the computational complexity, particularly for the quartic
interaction terms. Typically $10^3$ Fourier grid points are needed for
good convergence.  Using the Message Passing Interface (MPI), a
parallel implementation with data partition is made.  Good linear
speed-up is obtained upto 64 processes.

In the mean-field treatment of the heat transport, there is a need to
find the variational derivatives of the Green's functions and
self-energy for the conductance calculation.  This is done by solving
iteratively the set of equations
\begin{eqnarray}
\!\!\!\!\delta G^r &\!\!=\!\!& G^r \delta \Sigma^r_n G^r, \\
\!\!\!\!\delta G^{<} &\!\!=\!\!&  G^r \delta \Sigma_n^r G^{<}\! +\! 
                G^r\!\left( \delta \Sigma^{<}\! +\! \delta \Sigma_n^{<} \right)\! G^a 
               \!+\! G^{<} \delta \Sigma_n^a G^{a}.
\end{eqnarray}
The expressions for $\delta \Sigma_n^r$ and $\delta \Sigma_n^{<}$ can
be obtained by differentiating the self-energy results directly.  For
convergence control, linear mixing of the new and old values is used.

\section{Applications}

\subsection{One-dimensional cubic onsite model}

One important question to ask about the nonequilibrium formulation is
that whether the theory gives ballistic and diffusive heat transport
in a single framework.  Clearly, if the nonlinear self-energy
$\Sigma_n$ can be computed accurately, there is no doubt that the
answer is yes.  However, this is not clear if only the leading order
perturbation theory is used.  In Ref.~\cite{spohn}, Spohn gives a
derivation of a Boltzmann equation for the phonons in a simple 1D
chain with onsite cubic nonlinearity.  It is clear that in the
Boltzmann equation framework, we obtain Fourier's law and diffusive
heat transport.  We demonstrate in this subsection that diffusive heat
transport is not observed if only perturbative result is used. This
subsection also serves as an example of illustration of the concepts
discussed earlier by a simple and concrete model.

We consider a 1D chain with inter-particle spring constant $K$ and
onsite spring constant $K_0$.  We take a simple cubic onsite potential
for the interaction, $\frac{1}{3} t \sum_{j} u_j^3$, for $j$ belonging
to the central region.  This means that $T_{ijk} = t \delta_{ij}
\delta_{ik}$.  Apart from the nonlinear interaction, the left lead,
central region, and the right lead are identical.  The classical
equation of motion is
\begin{equation}
\ddot{u}_j = K u_{j-1} + (-2K-K_0) u_j + K u_{j+1} - t_j u_j^2,
\end{equation}
where $t_j = t$  for $ 1 \leq j \leq N$ and $t_j = 0$ for
$j < 1$ or $j > N$.  

The retarded Green's function $G_0^r$ can be obtained by solving the
linear equation $[ (\omega + i \eta)^2 - \tilde K] G_0^r = I$, where
(infinite in both directions) matrix $\tilde K$ is $2K+K_0$ on the
diagonal and $-K$ on the first off-diagonals.  The explicit solution
is
\begin{equation}
{G_0}^r_{jl}[\omega] = \frac{ \lambda_1^{|j-l|} }{(\lambda_1 - \lambda_2) K},
\end{equation}
where $\lambda_1$ and $\lambda_2$ are the roots of the equation
\begin{equation}
K \lambda^{-1} + \Omega  + K \lambda = 0,\quad
\Omega = (\omega + i \eta)^2 - 2K - K_0 .
\end{equation}
Note that $\lambda_1 \lambda_2 = 1$.  We define the two roots such
that $|\lambda_1| < 1$ and $|\lambda_2| > 1$.

The surface Green's function $g^r$ satisfies a similar equation as
$G_0^r$ except that it is semi-infinite in extent.  We consider the
left lead ($j \leq 0$).  The result for the right lead is identical.
Since the matrix $V^{LC}$ is nonzero only for one corner element, we
need the $g_0 = g_{00}$ component of the matrix $g^r$.  Consider only
the $j=0$ column, the equations for $g^r$ in component form are
\begin{equation}
\Omega g_0 + K g_{-1} = 1,
\end{equation}
\begin{equation}
 K g_{j-1} + \Omega g_j + K g_{j+1} = 0,\quad j=-1, -2, \cdots 
\end{equation}
Substituting the trial solution $g_j = c \lambda_1^{-j}$, we find
that 
\begin{equation}
g_0 = - \frac{\lambda_1}{K}.
\end{equation}
The ``less than'' Green's function for the left lead is then
(c.f. Eq.~(\ref{GrtoGl}))
\begin{equation}
g^{<}_L = 2 i f_L {\rm Im}\, g_0 = - 2 i f_L \frac{{\rm Im}\, \lambda_1}{K},
\end{equation}
where $f_L$ is the Bose-Einstein distribution function at the
temperature of the left lead.  The $G_0^{<}$ Green's function can be
obtained with the equation $G_0^{<} = G_0^r \Sigma^{<} G_0^a$, $\Sigma
= \Sigma_L + \Sigma_R$; $\Sigma_L = V^{CL} g_L V^{LC}$, similarly for
$\Sigma_R$.  The matrix elements of $\Sigma^{<}$ are all zero except
that $\Sigma_{11} = K^2 g^{<}_L$ and $\Sigma_{NN} = K^2 g^{<}_R$.
Using these results, after some calculation, we can write
\begin{equation}
{G_0}^{<}_{jl}[\omega] = 
\frac{ f_L \lambda_1^{j-l} + f_R \lambda_1^{l-j} }%
{(\lambda_1 - \lambda_2 ) K} \Theta(\omega).
\end{equation}
The $\Theta(\omega)$ function is defined to be 1 inside the phonon
band and zero outside the band; more precisely, $\Theta(\omega) = 1$
if $ K_0 < \omega^2 < 4K + K_0$, and is 0 otherwise.

Due to the onsite cubic nonlinearity, the leading order nonlinear
self-energy calculation is simple.  The first graph contribution is
\begin{equation}
{}^{1}{\Sigma_{n}}^{\sigma\sigma'}_{jl}(t) = 2 i\, t^2 
\bigl[{G_0}^{\sigma\sigma'}_{jl}(t)\bigr]^2.
\end{equation}
The contribution from the second graph is a constant in $\omega$:
\begin{equation}
{}^{2}{\Sigma_{n}}^{\sigma\sigma'}_{jl}[\omega]  = 2 i\, t^2 
\sigma \delta_{\sigma, \sigma'} \delta_{jl} \sum_{\sigma'',s}
\sigma'' {G_0}^{\sigma\sigma''}_{js}[0]
{G_0}^{\sigma''\sigma''}_{ss}(0).
\end{equation}
Both the summation over $\sigma''$ and over index $s$ can be performed
analytically.  We find
\begin{equation}
{}^{2}{\Sigma_{n}}^{\sigma\sigma'}_{jl}[\omega]  =  2 i\, t^2 
\sigma \delta_{\sigma, \sigma'} \delta_{jl} \bar{G}_j^r[0] \bar{G}^t(0),
\end{equation}
where 
\begin{eqnarray}
\bar{G}_j^r[\omega] &=& 
\frac{ 1 + \lambda_1 - \lambda_1^j - \lambda_1^{N-j+1} }%
{(1-\lambda_1)(\lambda_1 - \lambda_2) K},\\
\bar{G}^t[\omega] &=& \frac{ 1 + (f_L + f_R) \Theta(\omega) }%
{(\lambda_1 - \lambda_2) K}, \\
\bar{G}^t(0) &=& \frac{1}{2\pi} \int_{-\infty}^{+\infty} \!\!\!\!
\bar{G}^t[\omega] d\omega.
\end{eqnarray}

For numerical accuracy, we compute the conductance directly instead of
using finite differences. Let the temperature on the left lead be $T_L
= T + \Delta T/2$ and right lead be $T_R = T - \Delta T/2$.  We define
the thermal conductance as that in Eq.~(\ref{conductance-def}) and the
effective transmission by Eq.~(\ref{conductance-2}).  For the
expression for effective transmission $\tilde{T}[\omega]$, we also
need to take the limit $\Delta T \to 0$.  The advantage of doing this
is that the leading terms corresponding to equilibrium Green's
functions cancel exactly, since there should be no heat current when
$T_L = T_R$.  We can also equivalently define the temperature
variation through the expansion:
\begin{equation}
G^r = G^r_{{\rm eq}} + \frac{\delta G^r}{\delta T} \Delta T + O(\Delta T^2),
\label{vari}
\end{equation}
and similarly for $G^<$.  We can simplify the effective transmission
as
\begin{eqnarray}
\tilde{T}[\omega] &=& \frac{1}{2} {\rm Tr}\Bigl\{
(G^r - G^a)(\Sigma_R^{<} - \Sigma_L^{<}) + \nonumber\\
&&\quad\quad\quad\quad i G^{<} (\Gamma_R - \Gamma_L)
\Bigr\}/(f_L - f_R) \nonumber \\
&=& i K ({\rm Im}\, \lambda_1) ( F_{11}^L - F_{NN}^R ),
\end{eqnarray}
where the $F$ function is defined as
\begin{equation}
\label{F-vari}
F^L = \frac{1}{2} (G^r - G^a) + \Bigl[ f \frac{\delta (G^r - G^a)}{\delta T} - 
\frac{\delta G^{<}}{\delta T} \Bigr]
\left(\frac{\partial f}{\partial T}\right)^{-1}.
\end{equation}
$F^R$ is the same except that the first term is negative.  In
evaluating the $F$ functions, we should set $T_L = T_R = T$ as we have
already taken the $\Delta T \to 0$ limit.  The variations required in
Eq.~(\ref{F-vari}) can be computed by the variation in nonlinear
self-energy, which in turn can be computed by the variation in the
linear Green's function $G_0$.  Due to the Dyson equation, the
variation of the retarded Green's function has a simple result:
\begin{equation}
\frac{\delta G^r}{\delta T } = 
G^r \frac{\delta \Sigma_n^r }{\delta T} G^r.
\end{equation}
The variation in $G^{<}$ is rather complicated:
\begin{eqnarray}
\delta G^{<} &=& \frac{\delta G^{<}}{\delta T} \Delta T 
=  G^r \bigl\{ \Sigma_n^r \delta G_0^{<} + 
\nonumber \\
&& \delta \Sigma_n^r \left[ \bar{G}_0^{<} + G^r ( \Sigma_n^{<} G^a +
\Sigma_n^r \bar{G}_0^{<} ) \right] \bigr\} - {\rm h.c.} 
\nonumber \\
& & +\, \delta G_0^{<} + G^r ( \delta \Sigma_n^{<} + 
\Sigma_n^r \delta G_0^{<} \Sigma_n^a) G^a,
\end{eqnarray}
where $\bar{G}_0^{<} = G_0^{<}(I + \Sigma_n^a G^a)$, h.c. stands for the
Hermitian conjugate of the preceding term, and $\delta G_0^r = 0$,
\begin{eqnarray}
\frac{\delta {G_0}^{<}_{jl}}{\delta T} &=& 
\frac{\lambda_1^{j-l} - \lambda_1^{l-j}}{2(\lambda_1 - \lambda_2)K} 
\Theta(\omega) \frac{\partial f}{\partial T},\\ 
\delta {\Sigma_n}^{r,<}_{jl}[\omega] &=& \frac{4i\, t^2}{2\pi} \!\!
\int_{-\infty}^{+\infty} \!\!\!\!\! d\omega'\, {G_0}^{r,<}_{jl}[\omega-\omega']
\delta {G_0}^{<}_{jl}[\omega']. \quad
\end{eqnarray}

\begin{figure}
\includegraphics[width=\columnwidth]{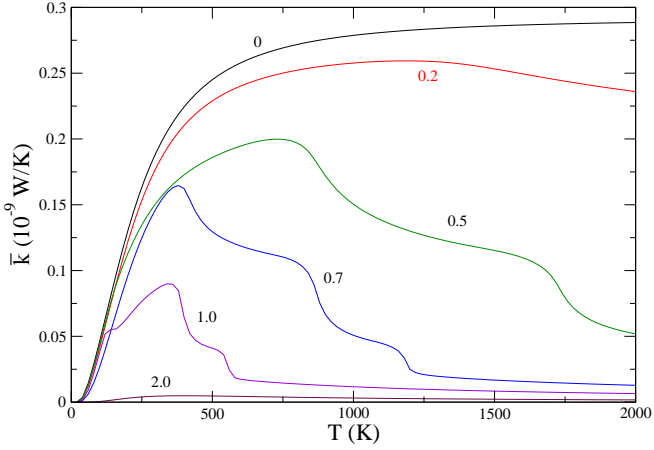}
\caption{\label{kappa}Thermal conductance of the cubic onsite model as
a function of temperature for several values of the cubic coupling
strength $t=0$, 0.2, 0.5, 0.7, 1.0, and 2.0 eV/(\AA$^3$u$^{3/2}$).
The length of center chain is $N=5$.}
\end{figure}
\begin{figure}
\includegraphics[width=\columnwidth]{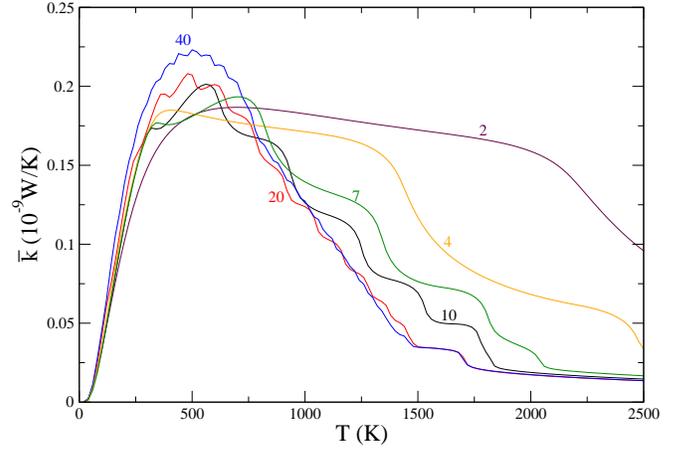}
\caption{\label{size-datB}Thermal conductance as a function of
temperature for several system lengths $N$ with $t=0.5$
eV/(\AA$^3$u$^{3/2}$).}
\end{figure}

We have made analytic calculations as far as possible.  The rest of
integrations and matrix inversions are calculated numerically.  We use
physical units with $K=0.625\,$eV/(\AA$^2$u), $K_0 = 0.1\,K$, (u for
unified atomic mass units). Figure~\ref{kappa} shows the temperature
dependence of the thermal conductance $\bar{\kappa}$ of a short
nonlinear chain of length $N=5$.  The $t=0$ curve is the ballistic
Landauer formula result, the rest of the curves show the effect of
nonlinearity.  The prominent feature here is that the conductance
decreases quickly as the nonlinear coupling $t$ increases from 0 to 2.
In particular, the system behaves like a thermal insulator if the
on-site nonlinearity is large enough.  The thermal conductance is
small at low temperatures and at high temperatures, as was found
experimentally.  At the extremely high temperatures, the conductance
decreases with temperature according to $1/T$ (data not shown).  This
agrees with several other theories of thermal conductance at high
temperatures.

Can we see diffusive transport in this model (under the perturbative
approximation)?  The answer is no.  In Fig.~\ref{size-datB}, we show
the size dependence for a fixed $t$.  At temperatures below 200$\,$K,
the results are essentially independent of the lengths, showing
ballistic heat transport.  At around 2000$\,$K, from sizes 2 to 10, we
see decrease of the thermal conductance roughly as $1/N$, but this
trend does not continue, and the conductance saturates to some value
for systems larger than 20.  This feature is generic; the FPU model
with first-order perturbation treatment also has a similar behavior.
The curves also display some oscillatory structure.  The number of
oscillations appears to be roughly equal to half of the chain length
$N$.  The feature is damped out as the chain becomes longer.  It would
be nice if a molecular dynamics simulation could confirm these
features.

The failure to see diffusive behavior may be understandable as our
result takes into account only the first-order perturbation in the
nonlinear interaction.  Other features seem qualitatively correct in
comparison with real systems; this is very encouraging.

\subsection{Fermi-Pasta-Ulam model}

The Fermi-Pasta-Ulam (FPU) model has been studied as a prototype model
in nonlinear dynamics \cite{FPU-model}.  In recent years, a lot of
works have been done on the thermal transport in the FPU model
\cite{lepri,libaowen}.  By molecular dynamics and mode-coupling
theory, Lepri {\sl et al.} \cite{2/5,lepri-mode-coupling} found that
the thermal transport in such a system does not have an ordinary
diffusive behavior in the sense that the thermal conductivity diverges
with the system size as some power, $L^{\alpha}$.  The actual value of
the exponent $\alpha$ is somewhat controversial.  The earlier studies
gave $2/5$, but the latest analysis suggests $1/3$ \cite{1/3,delfini}.

Quantum thermal transport in FPU model has seldom been studied.  In
particular, the change from ballistic to diffusive transport has not
been addressed.  This question can not be correctly answered within
molecular dynamics approach as the ballistic transport is
quantum-mechanical in nature.

We use realistic model parameters derived from expanding the Morse
potential, $V(x) = D \bigl( e^{-a x} - 1 \bigr)^2$ with $D = 3.8\,$eV,
$a = 1.88\,$\AA$^{-1}$, up to the fourth order in $x = (u_i -
u_{i+1})/\sqrt{m}$, where $u_i$ is mass-renormalized displacement for
the $i$th particle.  It is a general FPU model with both cubic and
quartic interactions.  The leads do not have nonlinear interactions.
We use the mass of carbon, $m = 12\,$u, in the calculation.  Thus we
can consider the results correspond to that of a carbon chain.

\begin{figure}
\includegraphics[width=\columnwidth]{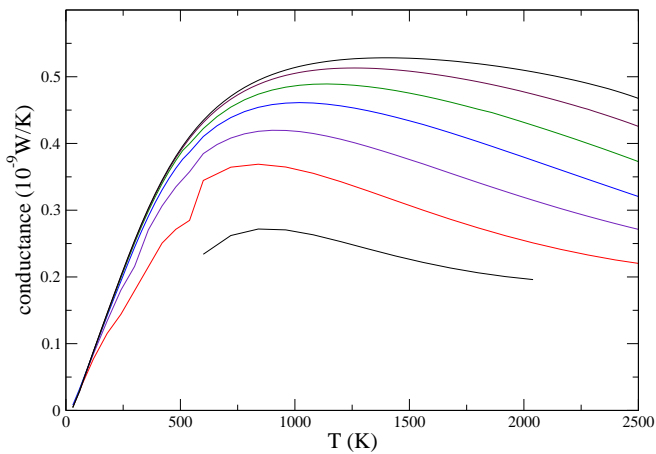}
\caption{\label{fig-Dexp}Thermal conductance of a Fermi-Pasta-Ulam
chain with chain lengths 4, 8, 16, 32, 64, 128, 256, from top to
bottom.  Semi-infinite one-dimensional harmonic chains serve as heat
baths with spring constants $k_L = 2.31$ and $k_R = 2.25$ eV/(u\AA$^2$)
for the left and right leads, respectively.  Small onsite quadratic
potentials are applied with $k_L^{\rm onsite} = 0.006$ and $k_R^{\rm
onsite} = 0.01$ eV/(u\AA$^2$).}
\end{figure}
 
In Fig.~\ref{fig-Dexp} we present the mean-field results of
conductance $\bar{\kappa}$ as a function of temperature for different
lengths of the chain.  As we can see, below about $300\,$K, the
conductance is independent of chain lengths; this is the ballistic
region of heat conduction.  At high temperatures, the conductance
decreases with the system size.  For a completely diffusive heat
transport, we should have $\bar{\kappa} \propto 1/L$.  We are far from
this.  Since in FPU model real diffusive behavior is impossible, the
reduction in conductance in the classical regime should be $L^{\alpha
-1} \approx L^{-0.6}$.  The observed exponent for sizes up to 256 is
much smaller.  In fact, this result is expected because the mean free
path in a typical carbon systems (such as carbon nanotubes
\cite{mean-free-path} or diamond) is of order $\mu\,$m.  Since the
conductivity decreases with temperature typically as $1/T$, even at
few thousand kelvin, the mean free path is expected to be 100 to 1000
lattice spacings.  Thus we see only tendency towards diffusive
behavior.  It is also clear that ballistic transport and diffusive
transport are at quite different time and length scales.  The present
calculation is already very large on an IBM 500-processor parallel
computer. It is out of the possibility to see diffusive behavior for
large sizes within the present approach.  But this example does give
us an indication that the mean-field theory could give diffusive
behavior for appropriate systems.

\begin{figure}
\includegraphics[width=\columnwidth]{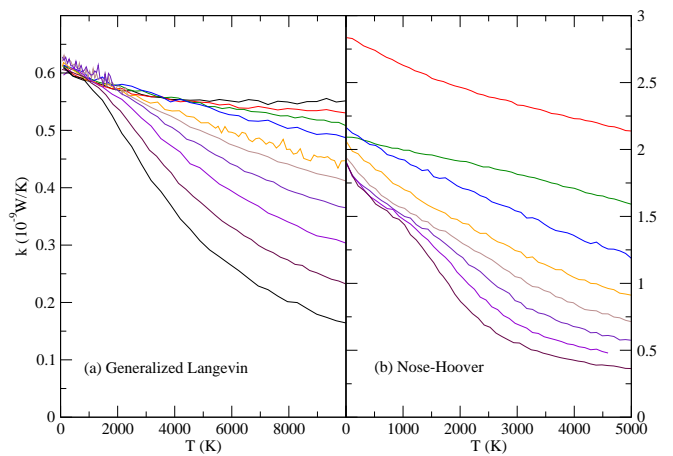}
\caption{\label{fig-md}Molecular dynamics results of the thermal
conductance for the model as in Fig.~\protect\ref{fig-Dexp}.  (a)
``Generalized Langevin'' thermal heat baths from sizes 4, 8, 16,
\dots, 2048, from top to bottom.  (b) Nos\'e-Hoover thermostat with
size 8, 16, 32, \dots, 1024.}
\end{figure}

It is instructive to compare the mean-field results with molecular
dynamics results.  In Fig.~\ref{fig-md}, we present the MD results
with the generalized Langevin-like heat-bath and the Nos\'e-Hoover
heat bath.  The colored noise is implemented using a standard spectrum
method with fast Fourier transform \cite{color-noise-book}.  It turns
out that the results for the thermal conductance is very sensitive to
the bath used and detail couplings between the bath and system.  It is
clear that quantum nature shows up already as high as $1000\,$K for
this model chain.  This can be understood by a very high Debye
temperature $T_D = \hbar \omega_{\rm max}/k_B \approx 2000\,{\rm K}$,
where $\omega_{\rm max} = \sqrt{4 K_c}$.  The Landauer formula,
Eq.~(\ref{conductance-2}), provides a rigorous upper bound for the
thermal conductance (when $\tilde T[\omega] = 1$, $T \to \infty$),
$\bar{\kappa}_{\rm max} = \frac{1}{2\pi} k_B \omega_{\rm max} \approx
0.65$ nW/K.  Interestingly, this is satisfied at low temperatures by
the generalized Langevin data.

To what extent the mean-field theory is ``correct''?  We can see that
it is qualitatively correct for the accessible temperature and size
ranges.  However, if we trust the Langevin heat-bath result, then, it
appears that the mean-field approximation somehow over estimates the
effect of nonlinearity.

\subsection{Benzene ring}

\begin{figure}
\includegraphics[width=\columnwidth]{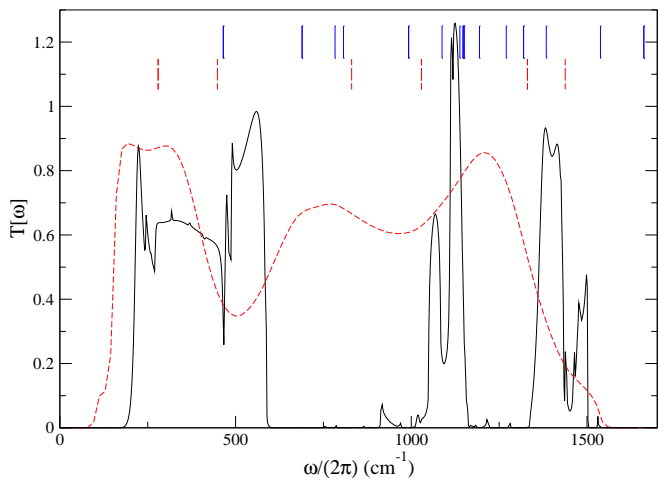}
\caption{\label{fig:benzenetr}Effective transmission of the benzene ring
junction at a temperature of $300\,$K. The solid line is for a full model
with force constants calculated from Gaussian, while the dotted line
is from a simplified 2D model.  The vertical bars indicate the
vibrational frequencies of isolated benzene rings of the full model
(topmost) and simplified model (dashed), respectively.}
\end{figure}

Molecular devices have attracted a lot of attentions in recent years,
due to their potential for future generation electronics.  Experiments
have been conducted on such systems; theoretical analyses of their
electronic and thermal transport properties have been performed
\cite{galperin,ciraci-EPL}.

In this subsection, we examine the thermal transport properties of a
benzene molecule connected to two polyethylene leads.  Two opposite
sites of a benzene ring, which are normally the hydrogen atoms, are
now bound covalently to carbon atoms, forming a 1,4-dibutyl-benzene
molecule representing the whole system.  The system has 14 carbon
atoms and 22 hydrogen atoms.  The interactions are calculated using
Gaussian 03 \cite{g03}.  We use the Hartree-Fock method with the 6-31G
basis set.  We have also used a convenient new feature of Gaussian,
which can calculate anharmonic interactions up to the 4th order.

Fig.~\ref{fig:benzenetr} (solid line) shows the effective transmission
function $\tilde{T}[\omega]$ of the benzene junction model at a
temperature of $300\,$K calculated using the mean-field theory.  The
peak features are complicated, resulting from the vibrational modes of
the molecule, the interaction with the leads, and the vibrational
spectra of the leads, as well as the temperature effect.  To
understand the transmission function better, we correlate the peak
features with the vibrational spectrum of an isolated benzene molecule
calculated from Gaussian, shown in Fig.~\ref{fig:benzenetr} as solid
bars.  The transmission peaks are roughly located at where the
vibrational frequencies appear, but somewhat shifted to low
frequencies.  It is difficult to identify the peaks with eigenmodes
unambiguously.

We notice that the hydrogen atoms do not play much role in the thermal
transport.  In particular, there are frequencies around $3000\,{\rm
cm}^{-1}$ associated with the vibrations of H atoms, but there is no
transmission in these high frequencies.  Vibrations of H atoms have
only local effects.  We confirm that the H atoms are less important by
a much simplified model retaining only the degrees of freedom of the
carbon atoms in a 2D model.  The leads are pseudo polyethylene chains;
we use an united atom approximation, so the CH$_2$ group is treated as
a single atom. The force constant between each atom is obtained from
the DREIDING model \cite{Mayo:1990}. There are only bond interactions
in the leads, and we have used the Morse function to calculate the
bond potential, which has the form $D(e^{-\alpha(R-R_0)}-1)^2$, where
$D=70\,$(kcal/mol), $\alpha=\sqrt{5}\,\mbox{\AA}^{-1}$, and $R-R_0$ is
the displacement (with the equilibrium bond length $R_0 = 1.4\,$\AA).
For the benzene ring, we also include an angular potential, which has
the form $\frac{4}{3}\bigl(\cos(\theta_{ijk})+\frac{1}{2}\bigr)E$,
where $E=100$ (kcal/mol).  We make a Taylor expansion of the potential
and calculate the force constants up to the 4th order.

The transmission for the simplified model (dotted line in
Fig.~\ref{fig:benzenetr}) is smoother due to much reduced degrees of
freedom.  It is qualitatively in agreement with the full model.  In
this case, the peaks appear to be in good agreement with the
vibrational modes of a hexagonal ring.

\subsection{Carbon nanotube junction}

\begin{figure}
\includegraphics[width=\columnwidth]{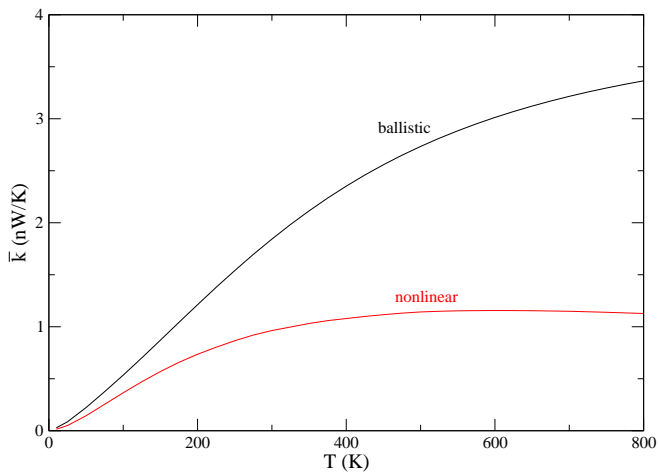}
\caption{\label{fig-ntb}Thermal conductance $\bar{\kappa}$ as a
function of temperature $T$ for a twenty-atom (5,0) carbon nanotube
junction.}
\end{figure}

Thermal transport of carbon nanotubes has been studied by molecular
dynamics \cite{berber,maruyama,yao,g-zhang}, Landauer formula
ballistic transport \cite{yamamoto}, and a phenomenological theory
taking into account the nonlinear effect \cite{wang-jian}.  There are
also experimental works on carbon nanotubes \cite{exp-papers}.  We
have also studied the heat transport properties of the $(5,0)$ carbon
nanotube using a tight-binding (TB) method \cite{NRL-TB}, of which the
TB parameters for the carbon atom have been checked recently
\cite{Fyta}. The left lead, the central junction, and the right lead
consist of forty, twenty, and forty atoms, respectively.  A unit cell
of the $(5,0)$ nanotube has twenty atoms.  The tube's axial dimension
is long enough (about $20\,$\AA) to allow a single $\Gamma$ point to
be used for the $k$-point sampling. We have written a force subroutine
in order to perform atomic relaxation of the nanotube as well as to
calculate force constants by numerical finite differences.  The spring
constants $K^{\alpha}$, the nonlinear force constants $T_{ijk}$ and
$T_{ijkl}$ are systematically obtained by displacing one, two, and
three atoms, respectively. For each displacement, the forces acting on
the atoms are used in the central-difference scheme to calculate the
force constants. Since the TB method we have adopted is free from
numerical noises, we have used a reasonably small maximum displacement
of $h_{\rm max}$ of $3 \times 10^{-4}$~\AA\ (i.e., the atoms are
displaced not more than $h_{\rm max}$ from their equilibrium
positions).  We note that this value of $h_{\rm max}$ is two orders of
magnitude smaller than that commonly used in a supercell
force-constant method \cite{Gan06,Kawazoe}.

In Fig.~\ref{fig-ntb} we present the results of thermal conductance
calculations.  We find that for such a complicated system, the
mean-field iterations are hardly converging, thus we only give the
ballistic (where the nonlinear interactions are turned off), and a
first-order perturbative result using the (single-line) diagrams in
Fig.~\ref{fig-mf}.  It is clear that the nonlinear interactions have a
large effect even at moderately high temperatures.  The thermal
conductance is slightly peaked around $600\,$K for the nonlinear
result.  We expect that the result should be reliable at least up to
room temperatures.

\section{Conclusion} 

We presented a nonequilibrium Green's function method for thermal
transport in nano-junctions.  We gave a unique derivation of the
equations of motion of Green's functions and rules for generating and
computing Feynman diagrams in a contour-ordered Green's function
setting.  The nonlinear interaction was treated with a systematic
perturbative expansion for self-energies.  For practical calculations,
the mean-field approximation has to be adopted.  We are still left with a
technical difficulty that the mean-field equations sometimes do not
converge.  Several versions of computer programs have been written for
consistency checks (a Mathematica notebook for an initial fast
prototyping, a Fortran program with MPI for very general models, and a
specialized program for the 1D cubic onsite model).  We have applied
the programs to a number of model systems.  Several sophisticated
`first-principles' methods were used in the force-constant
calculations for realistic models.  For the FPU model, we can
accurately predict the ballistic thermal transport and crossover to
diffusive behavior at sufficiently high temperatures.  We proposed to
use generalized Langevin heat baths, which are more appropriate for
nanojunctions.  For the benzene ring model, we observed interesting
behaviors in the transmission function related to the atomic
vibrational modes of the molecule.  For carbon nanotubes, it is clear
that nonlinear effect is quite strong even at low temperatures.

The Green's function discussed here is more fundamental than the
phonon distribution function in the Boltzmann equation.  The
connection of the present formulation with the quantum Boltzmann
equation or the phonon Boltzmann equation is something worth
pursuing. Formally, they are related by the Wigner transformation.  If
we can reduce the matrix of the Green's functions to vector of
distribution functions (in momentum space), potentially much larger
systems can be treated if we are willing to forsake some atomic
details.  Clearly, the nonequilibrium Green's function
method is at its best for small nanoscale systems.  A combination of the
present method with electronic transport and electron-phonon
interactions is worthy of further explorations.

\section*{Acknowledgements}

We thank Lin Yi for introducing us the nonequilibrium Green's function
method.  We also thank Hong Guo, Qinfeng Sun, Sai Kong Chin, and
Jingtao L\"u for discussions.  C. K. G. thanks
D. A. Papaconstantopoulos, L. Shi, and M. J. Mehl for a copy of the
tight-binding program {\it static} and various initial help with the
use of the program.  This work was supported in part by a National
University of Singapore Academic Research Fund.  The computations were
carried out on the Institute of High Performance Computing IBM
clusters and on the Singapore-MIT Alliance Opteron cluster.


\begin{thebibliography}{01}
\bibitem{peierls} R. E. Peierls, \textsl{Quantum Theory of Solids}
(Oxford University Press, 1955), Chap.~2.

\bibitem{heat-transport-textbooks} G. P. Srivastava,
\textsl{The Physics of Phonons} (Adam Hilger, Bristol, 1990).

\bibitem{rmp-review} P. Carruthers, Rev. Mod. Phys. \textbf{33},
92 (1961). 

\bibitem{cahill} D. G. Cahill, W. K. Ford, K. E. Goodson, G. D. Mahan,
A. Majumdar, H. J. Maris, R. Merlin, and S. R. Phillpot, J. Appl. Phys.
\textbf{93}, 793 (2003).

\bibitem{galperin} M. Galperin, M. A. Ratner, and A. Nitzan,
cond-mat/0612085.

\bibitem{landauer-thermal} L. G. C. Rego and G. Kirczenow, Phys. Rev. Lett. 
\textbf{81}, 232 (1998); M. P. Blencowe, Phys. Rev. B \textbf{59}, 4992 (1999).

\bibitem{michel} M. Michel, G. Mahler, and J. Gemmer,
Phys. Rev. Lett. \textbf{95}, 180602 (2005);
M. Michel, J. Gemmer, and G. Mahler, Int. J. Mod. Phys. B \textbf{20},
4855 (2006).


\bibitem{haanggi} D. Segal, A. Nitzan, and P. H\"anggi, J. Chem. Phys. 
\textbf{119}, 6840 (2003); 
D. Segal and A. Nitzan, Phys. Rev. Lett. \textbf{94}, 034301 (2005).

\bibitem{chenG} G. Chen, Phys. Rev. Lett. \textbf{86}, 2297 (2001).

\bibitem{canala} C. V. D. R. Anderson and K. K. Tamma, 
Phys. Rev. Lett. \textbf{96}, 184301 (2006). 

\bibitem{ciraci} A. Ozpineci and S. Ciraci, Phys. Rev. B \textbf{63},
125415 (2001).

\bibitem{mingo} N. Mingo and L. Yang, Phys. Rev. B \textbf{68},
245406 (2003). 

\bibitem{yamamoto} T. Yamamoto and K. Watanabe, Phys. Rev. Lett, 
\textbf{96} 255503 (2006).

\bibitem{dhar} A. Dhar and D. Sen, Phys. Rev. B \textbf{73}, 085119 (2006). 

\bibitem{PRB-green} J.-S. Wang, J. Wang, and N. Zeng, 
Phys. Rev. B \textbf{74}, 033408 (2006). 

\bibitem{Mingo-PRB-negf} N. Mingo, Phys. Rev. B \textbf{74}, 125402 (2006). 

\bibitem{meir-wingreen} Y. Meir and N. S. Wingreen,
Phys. Rev. Lett. \textbf{68}, 2512 (1992); 
A.-P. Jauho, N. S. Wingreen, and Y. Meir, Phys. Rev. B \textbf{50}, 5528 (1994). 
%
\bibitem{haug} H. Haug and A.-P. Jauho, \textsl{Quantum Kinetics in 
Transport and Optics of Semiconductors} (Springer, Berlin, 1996).

\bibitem{kadnoff-baymn} L. Kadanoff and G. Baymn, 
\textsl{Quantum Statistical Mechanics}
(W. A. Benjamin, New York, 1962).

\bibitem{doniach} S. Doniach and E. H. Sondheimer, 
\textsl{Green's Functions for Solid State Physicists}
(W. A. Benjamin, Reading, 1974), Chap.~1. 

\bibitem{mahan} G. D. Mahan,  
\textsl{Many Particle Physics}, 3rd Ed. (Springer, New York, 2000).

\bibitem{niu} C. Niu, D. L. Lin, and T.-H. Lin, J. Phys.: Condens. Matter,
\textbf{11}, 1511 (1999). 

\bibitem{keldysh} L. V. Keldysh, Sov. Phys. JETP, \textbf{20}, 1018 (1965). 

\bibitem{fetta-walacka} A. L. Fetter and J. D. Walecka, 
\textsl{Quantum Theory of Many-Particle Systems} (McGraw-Hill, New York, 1971).

\bibitem{valle} R. G. D. Valle and P. Procacci, Phys. Rev. B \textbf{46},
6141 (1992).

\bibitem{langreth} D. C. Langreth, in \textsl{Linear and Nonlinear
Electron Transport in Solids}, edited by J. T. Devreese and E. van Doren
(Plenum, New York, 1976), pp.3-32.

\bibitem{landauer} R. Landauer, IBM J. Res. Dev. \textbf{1}, 223 (1957);
Phil. Mag. \textbf{21}, 863 (1970).

\bibitem{caroli} C. Caroli, R. Combescot, P. Nozieres, and D. Saint-James,
J. Phys. C \textbf{4}, 916 (1971). 
 
\bibitem{Dhar-heat-bath} A. Dhar and D. Roy, 
J. Stat. Phys. 125, 801 (2006);
see also A. Dhar and K. Wagh, cond-mat/0604170.

%
\bibitem{G-Langevin} K. Lindenberg and B. J. West, 
\textsl{The Nonequilibrium Statistical Mechanics of Open
and Closed Systems} (VCH Publishers, New York, 1990), p.194.

\bibitem{surface-green} M. P. L\'opez Sancho, J. M. L\'opez Sancho,
and J. Rubio, J. Phys. F: Met. Phys. \textbf{15}, 851 (1985).

\bibitem{spohn} H. Spohn, math-ph/0505025;
K. Aoki, J. Lukkarinen, H. Spohn, J. Stat. Phys. \textbf{124}, 1105 (2006).

\bibitem{FPU-model} Focus issue: the ``Fermi-Pasta-Ulam'' problem -- the
first 50 years. Chaos, \textbf{15}, (2005). 

\bibitem{lepri} S. Lepri, R. Livi, and A. Politi, Phys. Rep. \textbf{377},
1 (2003). 

\bibitem{libaowen} B. Li, J. Wang, L. Wang, and G. Zhang, Chaos, \textbf{15},
015121 (2005).

\bibitem{2/5} S. Lepri, R. Livi, and A. Politi, 
Europhys. Lett. \textbf{43}, 271 (1998).

\bibitem{lepri-mode-coupling} S. Lepri, Phys. Rev. E \textbf{58}, 
7165 (1998).

\bibitem{1/3} O. Narayan and S. Ramaswamy, Phys. Rev. Lett.  \textbf{89}, 200601 (2002);
T. Mai, A. Dhar, O. Narayan, cond-mat/0608034.

\bibitem{delfini} L. Delfini, S. Lepri, R. Livi, and A. Politi,
Phys. Rev. E \textbf{73}, 060201(R) (2006); 
cond-mat/0611278.

\bibitem{mean-free-path} Y. Xiao, X. H. Yan, J. X. Cao, and J. W. Ding,
J. Phys: Condens. Matter, \textbf{15}, L341 (2003). 

\bibitem{color-noise-book} J. Garc\'ia-Ojalvo and J. M. Sancho,
\textsl{Noise in Spatially Extended Systems} (Springer-Verlag, New York,
1999).

\bibitem{ciraci-EPL} A. Buldum, D. M. Leitner, and S. Ciraci, 
Europhys. Lett. \textbf{47}, 208 (1999).

\bibitem{g03} M. J. Frisch, \textsl{et al.}, Gaussian 03, Revision C.02,
(Gaussian Inc., Wallingford, 2004).


\bibitem{Mayo:1990} S. L. Mayo, B. D. Olafson, and W. A. Goddard III,
J. Phys. Chem. \textbf{94}, 8897 (1990). 


\bibitem{berber} S. Berber, Y.-K. Kwon, and D. Tom\'anek, 
Phys. Rev. Lett. \textbf{84}, 4613 (2000). 

\bibitem{maruyama} S. Maruyama, Physica B \textbf{323}, 193 (2002).

\bibitem{yao}
Z. Yao, J.-S. Wang, B. Li, and G.-R. Liu, 
Phys. Rev. B \textbf{71}, 085417 (2005).

\bibitem{g-zhang} G. Zheng and B. Li, J. Chem. Phys. 
\textbf{123}, 114714 (2005).

\bibitem{wang-jian} J. Wang and J.-S. Wang, Appl. Phys. Lett. 
{\bf 88}, 111909 (2006).

\bibitem{exp-papers} P. Kim, L. Shi, A. Majumdar, and P. L. McEuen,
Phys. Rev. Lett. \textbf{87}, 215502, (2001);  
E. Pop, D. Mann, Q. Wang, K. Goodson, and H. Dai,
Nano Letters, \textbf{6}, 96 (2006).

\bibitem{NRL-TB} R. E. Cohen, M. J. Mehl, and D. A. Papaconstantopoulos, 
Phys. Rev. B {\bf 50}, 14694 (1994); 
M. J. Mehl and D. A. Papaconstantopoulos, Phys. Rev. B {\bf 54}, 4519 (1996); 
D. A. Papaconstantopoulos, M. J. Mehl, S. C. Erwin, and M. R. Pederson, 
Mater. Res. Soc. Proc. {\bf 491}, 221 (1998).

\bibitem{Fyta} M. G. Fyta, I. N. Remediakis, P. C. Kelires, and 
D. A. Papaconstantopoulos, Phys. Rev. Lett. {\bf 96}, 185503 (2006).

\bibitem{Gan06}
C. K. Gan, Y. P. Feng, and D. J. Srolovitz, Phys. Rev. B {\bf 73}, 235214 (2006).

\bibitem{Kawazoe}
K. Parlinski and Y. Kawazoe, Phys. Rev. B {\bf 60}, 15511 (1999).

\end{thebibliography}

\end{document}